\documentclass[%
 aip,
 amsmath,amssymb,
preprint,%
reprint,%
]{revtex4-1}

\usepackage{graphicx}
\usepackage{dcolumn}
\usepackage{bm}
\usepackage{xcolor}

\usepackage[utf8]{inputenc}
\usepackage[T1]{fontenc}
\usepackage{mathptmx}

\begin{document}

\preprint{AIP/123-QED}

\title[Temperature dependent elastic constants of BAs]{Temperature dependent elastic constants and thermodynamic
properties of BAs: an ab-initio investigation.}

\author{Cristiano Malica}
\email{cmalica@sissa.it}
 \affiliation{ 
International School for Advanced Studies (SISSA), Via Bonomea 265, 34136 Trieste (Italy).
}%

\author{Andrea Dal Corso}%
 \affiliation{ 
International School for Advanced Studies (SISSA), Via Bonomea 265, 34136 Trieste (Italy).
}%
\affiliation{%
IOM-CNR 34136 Trieste (Italy).
}%

\date{\today}

\begin{abstract}
We present an ab-initio study of the temperature dependent elastic
constants of boron arsenide (BAs), a semiconductor that exhibits ultra-high thermal
conductivity and is under investigation for thermal management in electronics.
We test the consistency of our predictions by computing the temperature dependent
sound velocity of the longitudinal acoustic mode along the $[111]$ direction and
comparing with experiment.
Furthermore, as a byproduct, we present the room temperature phonon dispersions, and
the temperature dependent thermal expansion, isobaric heat capacity, and
average Gr\"{u}neisen parameter comparing with the most updated experiments and
previous calculations when available. Finally, we present the theoretical estimate
of the temperature dependent mean square atomic displacements.
\end{abstract}

\maketitle

%

\section{Introduction}

Boron arsenide (BAs) is a promising semiconductor for its high thermal conductivity
at room temperature (RT) that makes it a possible candidate for applications
in electronics and photonics that require efficient heat dissipation from
hot spots in devices.
Although the growth of BAs was reported since 1950s, only recently high-quality
samples have been synthesized with the measurement
of an ultrahigh thermal conductivity up to 1300 $W m^{-1} K^{-1}$
~\cite{Kang_2017, Kang_2018, Li_2018, Tian_2018}.
These measurements spurred the interest for this material
so that many experiments and theoretical calculations have been carried
out to investigate its physical properties.
BAs has a zinc-blende cubic structure and belongs to the space group
$F\bar{4}3m$~\cite{Kang_2018, Li_2018, Tian_2018}. Recent measurements,
supported by ab-initio calculations, range
from the lattice constant and thermal expansion (TE), to the band gap and refractive
index, to the elastic constants (EC) and the bulk
modulus~\cite{Tian_2019, Kang_2019, Chen_2019}.

The EC are important quantities to describe the mechanical and thermodynamic
properties of materials since they allow the check of the crystal stability,
determine the speeds of sound, and can be used to compute the TE and
the thermal stresses. At $T=0$ K the ECs of BAs have been computed in several
works by means of density functional theory (DFT) using several exchange
and correlation functionals \cite{Wang_2003, Hassan_2004, Meradji_2004, 
Bing_2010, Daoud_2015, Tian_2019, Kang_2019}. However, in literature, the information
about the effect of temperature on the ECs is rather indirect. 
Kang {\it et al.}~\cite{Kang_2019}
reported the measured speed of sound of the
longitudinal acoustic mode along the $[111]$ direction from $T=80$ K to
$T=500$ K showing a decrease of about $2.4$ \%, but no theoretical calculation
is available to support this measurement and in general to estimate
the temperature dependent elastic constants (TDECs) of BAs.

We have recently implemented in the \texttt{thermo\_pw} code~\cite{Malica_2020}
the calculation of the TDECs both isothermal and adiabatic.
The TDECs can be computed from second derivatives of the Helmholtz
free-energy within the quasi-harmonic approximation (QHA) by means of
density functional theory (DFT) and density functional perturbation
theory (DFPT).
In this paper we apply the calculation of TDECs to BAs. We found that in the
range of temperature $0-1800$ K the percentage softening of adiabatic ECs is
$\approx$ 11 \% for $C_{11}$, $\approx$ 9 \% for $C_{12}$ and $\approx$ 13 \%
for $C_{44}$. In the range of temperature $0-800$ K the softening is comparable
but slightly smaller than the one of silicon.

As a byproduct of our calculation we report the RT
phonon dispersions of BAs comparing with the inelastic X-ray scattering
measurements and the temperature dependence of several other thermodynamic
quantities such as the TE, the isobaric heat capacity, the
average Gr\"uneisen parameter, and the atomic B-factors (BFs).
We compare these quantities with experimental data and previous calculations
when available. In general, the agreement is quite good.

\section{Theory}

The calculation of the TDECs within the QHA is explained in detail in our
recent work~\cite{Malica_2020}; in this section we limit ourselves to a summary
of the most important formulas in order to make the paper self-contained.

The isothermal ECs are obtained from the derivatives of the Helmholtz
free-energy $F$ with respect to strain $\epsilon$:
\begin{equation}\label{dUT}
\tilde C_{ijkl}^T = \frac{1}{\Omega} \left(\frac{\partial^2 F}{\partial \epsilon_{ij} 
\partial \epsilon_{kl}} \right)_{\epsilon=0}.
\end{equation}
Since we are usually interested in the ECs obtained from the stress-strain
relationship ($C_{ijkl}^T$) we correct the $\tilde C_{ijkl}^T$ when the
system is under a pressure $p$ as:
\begin{equation}\label{dUp}
C^T_{ijkl} = \tilde C^T_{ijkl} + \frac{1}{2} p \left(2 \delta_{ij} \delta_{kl} - \delta_{il} \delta_{jk} 
- \delta_{ik} \delta_{jl} \right).
\end{equation}
The Helmholtz free-energy of Eq.~\ref{dUT} is obtained as the sum of the DFT total
energy $U$ and the vibrational free energy (neglecting the electronic contribution):
$F=U+F_{vib}$.
The latter is given by:
\begin{eqnarray} \label{equ4}
F_{vib}(\mathbf \epsilon, T) =&&\frac{1}{2N} \sum_{\mathbf q \eta} \hbar \omega_{\eta} 
\left(\mathbf q, \mathbf \epsilon \right) + \nonumber\\
&&+ \frac{k_B T}{N} \sum_{\mathbf q \eta} \ln \left[1 - \exp \left(-\frac{\hbar \omega_{\eta}
(\mathbf q, \mathbf \epsilon)}{k_B T}\right) \right],
\end{eqnarray}
where $N$ is the number of cells in the crystal,
$\omega_{\eta}\left(\mathbf q, \epsilon \right)$ is the phonon angular frequency
of the mode $\eta$ with wave-vector $\mathbf q$ computed in the system with a
strain $\epsilon$.
Cubic solids have three independent ECs that in Voigt notations are $C_{11}$,
$C_{12}$, and $C_{44}$~\cite{nyebook}.
The QHA calculation of the ECs, Eq.~\ref{dUT}, is performed on a grid of reference
geometries by varying the lattice constant $a_0$.
Phonon dispersions are computed in the same grid in order to evaluate the free-energy
as a function of the volume, minimize it and obtain the temperature dependent crystal
parameter $a(T)$ ~\cite{Palumbo_2017, Palumbo_2017_2}. Then, at each temperature
$T$, the TDECs as a function of $a_0$ are interpolated and evaluated at $a(T)$.
The calculation requires phonon dispersions in all the strained configurations for
all the reference geometries, in addition to the phonon dispersions used
to compute $a(T)$.

The temperature dependent lattice parameter $a(T)$ can be used to derive the TE $\alpha$
as:
\begin{equation} \label{alfa}
\alpha = \frac{1}{a(T)} \frac{d a(T)}{d T}.
\end{equation}

The isochoric heat capacity within the harmonic approximation is given by:
\begin{equation} \label{cv}
C_{V} = \frac{k_B}{N} \sum_{\mathbf{q}\eta} \left(\frac{\hbar \omega_{\eta}
(\mathbf{q})}{k_B T}\right)^2  
\frac{ \exp( \hbar \omega_{\eta}(\mathbf{q})/k_B T) }
{ \left[ \exp( \hbar \omega_{\eta}(\mathbf{q})/k_B T) - 1 \right]^2 }.
\end{equation}
It is computed for each reference geometry and it is interpolated at $a(T)$.
The knowledge of $\alpha$ and $C_V$ allows to compute the
isobaric heat capacity $C_P$ and the average Gr\"{u}neisen parameter $\gamma$:
\begin{equation} \label{cp}
C_P = C_V + T \Omega \beta^2 B^T,
\end{equation}
\begin{equation} \label{gamma}
\gamma = \frac{\Omega \beta B^T}{C_V},
\end{equation}
where $\Omega$ is the volume of one unit cell at the temperature $T$, $\beta=3 \alpha$
is the volume TE and $B^T$ is the isothermal bulk modulus calculated from the
ECs as:
\begin{equation} \label{bmod}
B^T = \frac{1}{3} \left(C_{11}^T + 2 C_{12}^T \right).
\end{equation}
The adiabatic ECs $C_{ijkl}^S$ are obtained from the isothermal ones with
the relation:
\begin{equation} \label{adiab}
C_{ijkl}^S = C_{ijkl}^T + \frac{T \Omega b_{ij} b_{kl}}{C_{V}},
\end{equation}
where the $b_{ij}$ are the thermal stresses obtained from:
\begin{equation} \label{equ6}
b_{ij} = - \sum_{kl} C_{ijkl}^T \alpha_{kl}.
\end{equation}
The adiabatic bulk modulus $B^S$ is computed as in Eq.~\ref{bmod} in terms
of the adiabatic ECs.

The atomic BF is calculated as explained in a previous work~\cite{Malica_2019}.
Calling $B_{\alpha \beta}(s)$ the mean-square displacement matrix of the
atom $s$ we have:
\begin{equation} \label{bfact}
B_{\alpha \beta}(s) = \frac{\hbar}{2 N M_s} \sum_{\mathbf{q}\eta} \coth 
\left[ \frac{\hbar \ \omega_{\eta}(\mathbf{q})}{2 k_B T} \right]
\frac{u_{s \alpha}^\eta(\mathbf{q}) \left[ u_{s \beta}^\eta(\mathbf{q})\right]^*}
{\omega_{\eta}(\mathbf{q})},
\end{equation}
where $M_s$ is the mass of $s$-th atom, $u_{s \alpha}^{\eta}(\mathbf{q})$ is the
$s \alpha$-th component of the dynamical matrix eigenvector of the mode $\eta$
with wave-vector $\mathbf{q}$. The BF is defined as
$8 \pi^2 B_{\alpha \beta}(s)$ ~\cite{Malica_2019}.
In order to include anharmonic effects, BFs are computed in the same reference
geometries used for elastic constants and interpolated at each temperature
at the $a(T)$.

\begin{table*} \centering
\caption{The computed ECs compared with some results available in the literature. The exchange and correlation functionals are indicated
in the first column. The equilibrium lattice constant ($a_0$) is in \AA\ while the ECs and the bulk modulus are in kbar. The experimental
values are at T=298 K.}
\begin{tabular}{lccccc}
\hline
& \multicolumn{1}{c}{$a_0$} \ \ & \multicolumn{1}{c}{$C_{11}$} \ \ & \multicolumn{1}{c}{$C_{12}$} \ \ & \multicolumn{1}{c}{$C_{44}$} \ \ & \multicolumn{1}{c}{$B$} \\
\hline
\hline
LDA$^a$ & 4.745 & 2897 & 768 & 1557 & 1477\\
LDA$^b$ & 4.745 & 2897 & 768 & 1772 & 1477\\
LDA$^c$ & 4.756 & 2828 & 759 & 1520 & 1449\\
LDA$^d$ & 4.759 & 2807 & 754 & 1507 & 1438\\
LDA~\cite{Kang_2019} & 4.7444 & 2940 & 806 & 1770 & 1500\\
LDA~\cite{Daoud_2015} &       & 3013 & 772 & 1639 & 1519\\
LDA~\cite{Bing_2010} & 4.779 & 2864 & 710 & 1575 & 1428\\
LDA~\cite{Meradji_2004} & 4.743 & 2950 & 780 & 1770 & 1500\\
LDA~\cite{Wang_2003} & 4.721 & 2914 & 728 & 1579 & 1457\\
PBE~\cite{Tian_2019} & 4.817 & 2630 & 620 & 1430 & 1290\\
PBE~\cite{Meradji_2004} & 4.812 & 2750 & 630 & 1500 & 1340\\
PBE~\cite{Hassan_2004} & 4.784 & 2510 & 798 & 1270 & 1370\\
Expt.~\cite{Kang_2019} & 4.78 & 2850 & 795 & 1490 & 1480\\
\hline
\end{tabular}
\label{table1}
\\ $^a$ This work at $T=0$ K,
$^b$ This work at $T=0$ K with frozen ions,
$^c$ This work at $T=0$ K + ZPE,
\\ $^d$ This work at $T=300$ K (adiabatic ECs)
\end{table*}


\section{Method}

The calculations presented in this work were carried out using DFT as implemented
in the Quantum ESPRESSO package ~\cite{qe1, qe2}.
Thermodynamic properties have been computed using the \texttt{thermo\_pw}
package~\cite{tpw}.
The exchange and correlation functional was approximated by the local
density approximation (LDA) which gives the best agreement between experimental and theoretical
quantities especially for the $T=0$ K ECs. We employed the projector augmented wave (PAW) method and a plane waves basis
set with pseudopotentials~\cite{paw} from $pslibrary$~\cite{psl}.
The pseudopotentials \texttt{B.pz-n-kjpaw\_psl.1.0.0.UPF} and
\texttt{As.pz-n-kjpaw\_psl.1.0.0.UPF} have been used for boron and arsenic, respectively.
The wave functions (charge density) were expanded in a plane waves basis with a kinetic
energy cut-off of $60$ Ry ($400$ Ry), and a $16 \times 16 \times 16$ mesh
of \textbf{k}-points has been used for the Brillouin zone integration.
Density functional perturbation theory (DFPT)~\cite{rmp, dfptPAW} was used to calculate the
dynamical matrices on a $4 \times 4 \times 4$ \textbf{q}-points grid. The dynamical matrices have been
Fourier interpolated on a $200 \times 200 \times 200$ \textbf{q}-points mesh to evaluate the free-energy.
The grid of the reference geometries was centered at the $T = 0 \ K$ lattice constant reported
in Table~\ref{table1}. The number of reference geometries were 9 with lattice constants separated
from each other by $\Delta a = 0.05 \ a.u.$.
In the ECs calculation the number of strained configuration was 6 for each type of strain with an interval
of strains between two strained geometries of $0.005$. In total we computed phonon dispersions
for 162 geometries (in addition to the 9 phonon dispersions used for the $a(T)$ calculation).
A second degree polynomial has been used to fit the energy (for the ECs at $T = 0$ K) or
the free-energy
(for the QHA TDECs) as a function of strain, while a fourth degree polynomial was
used to fit at each temperature all the other quantities computed at the various
reference geometries versus $a(T)$.

\section{Applications}

\begin{figure*}
\centering
\includegraphics[width=.49\linewidth]{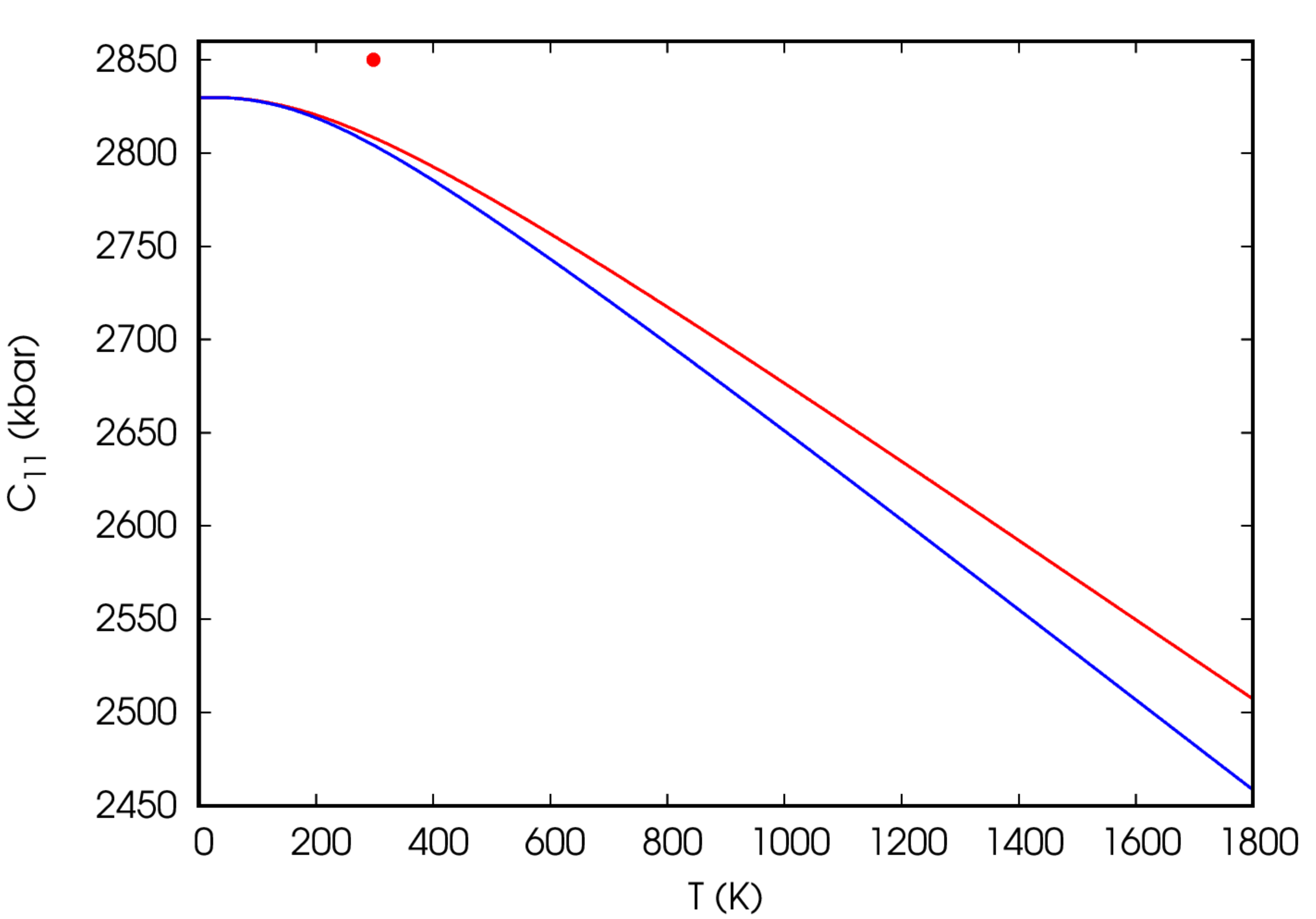}
\includegraphics[width=.49\linewidth]{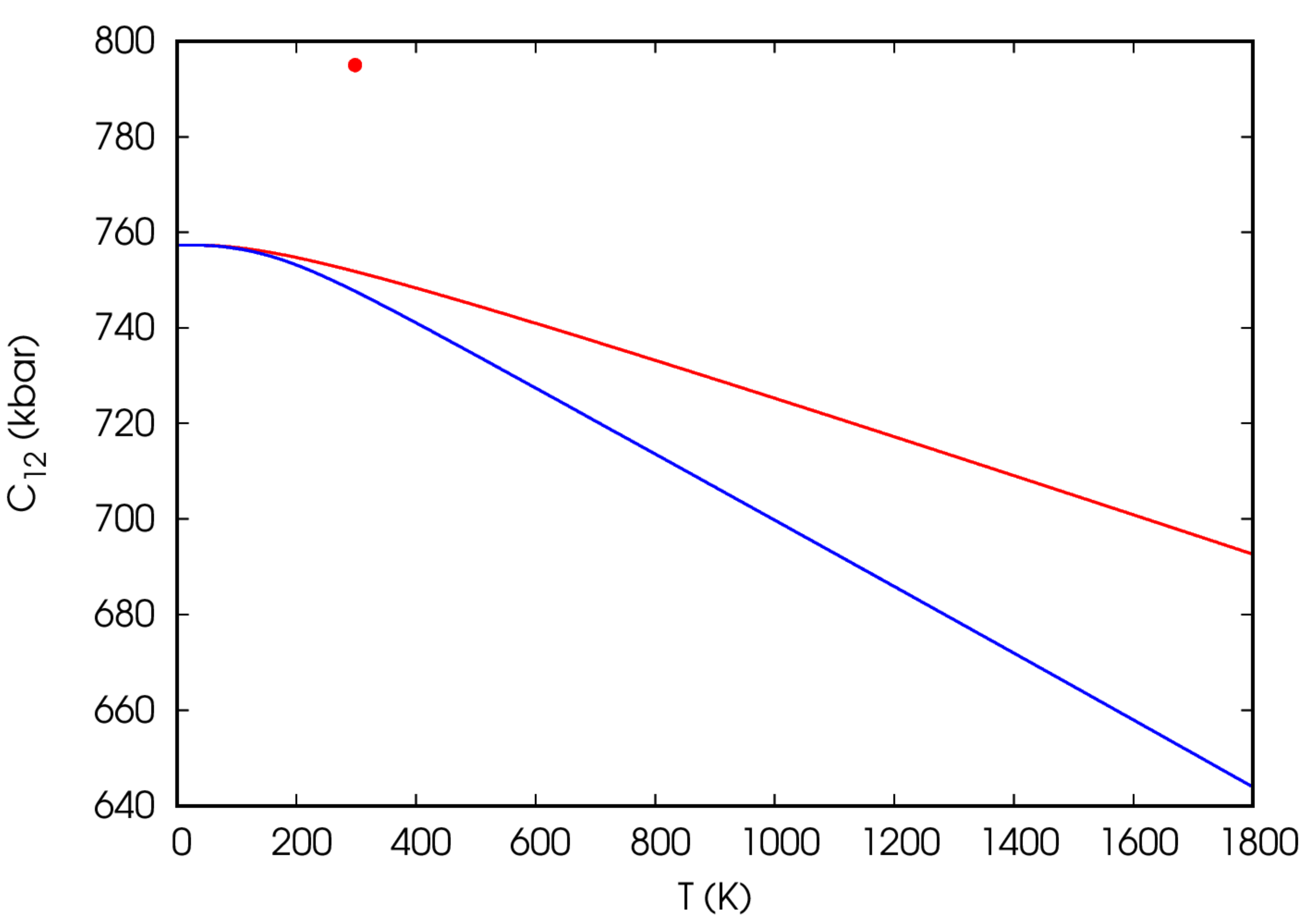}
\includegraphics[width=.49\linewidth]{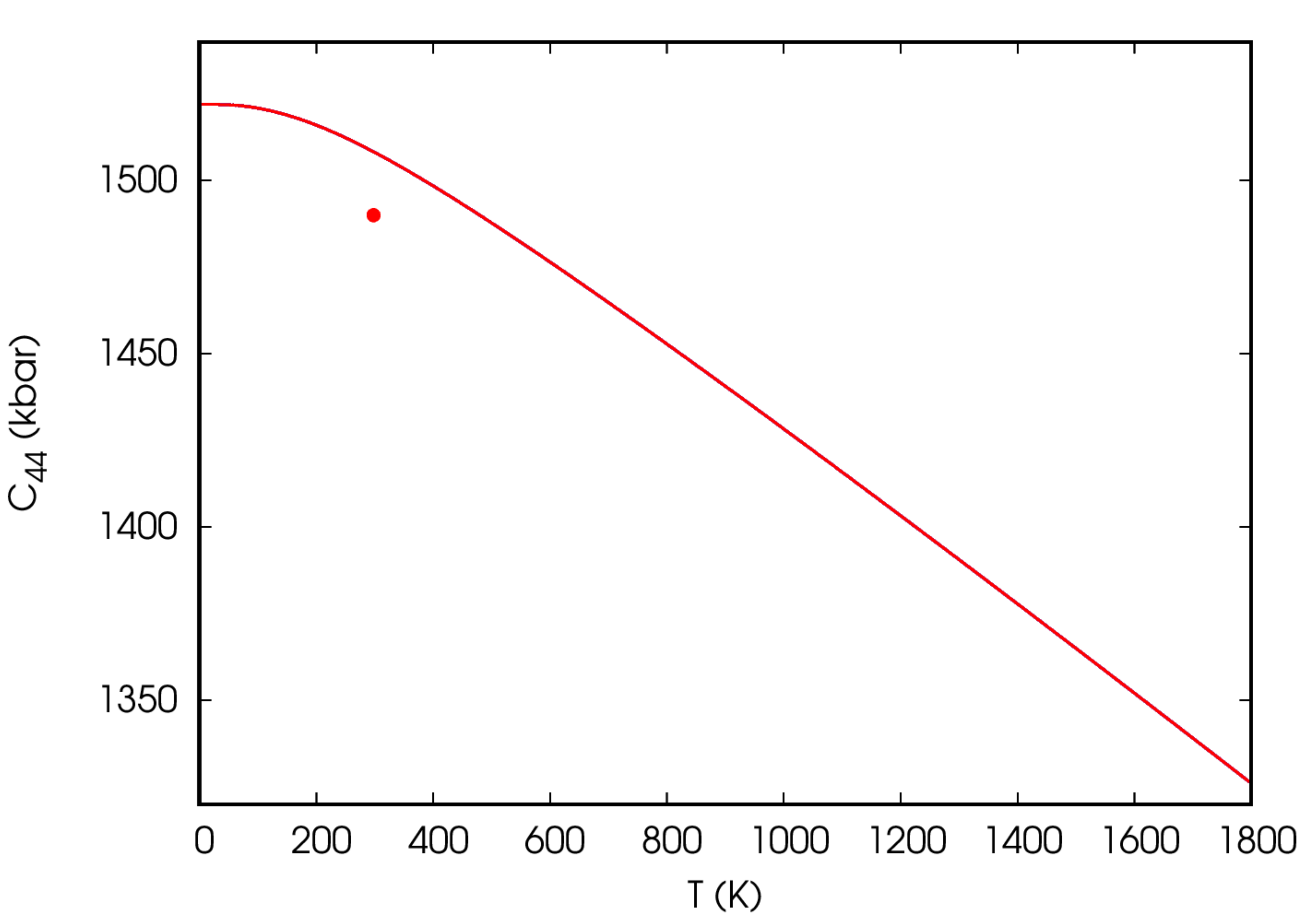}
\includegraphics[width=.49\linewidth]{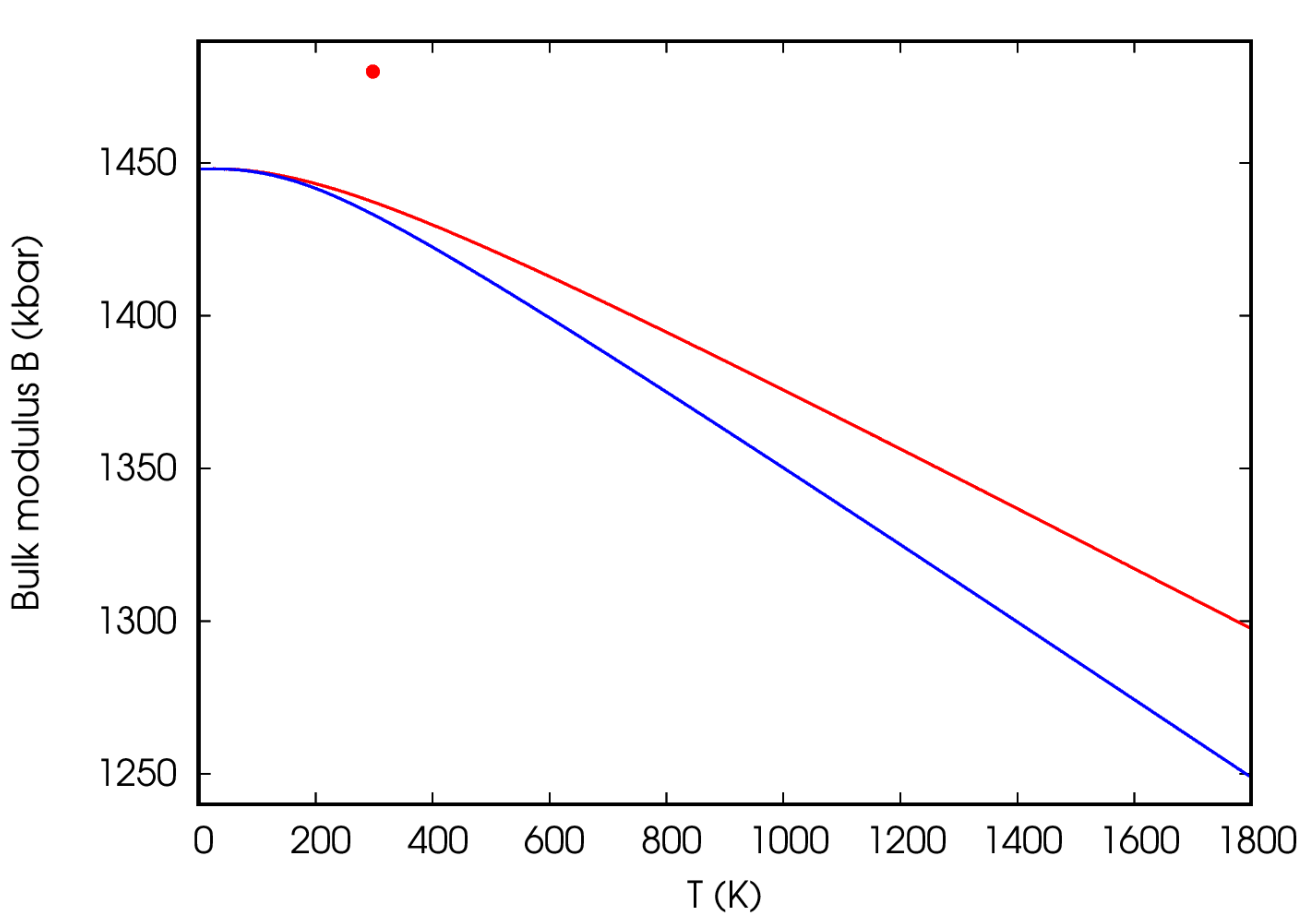}
\caption{TDECs and bulk modulus of BAs: isothermal (blue) and adiabatic (red). RT experimental points from Kang {\it et al.}~\cite{Kang_2019}.}
\label{fig_ec}
\end{figure*}

In Table~\ref{table1} we report the computed equilibrium lattice constant $a_0$
at $T=0$ K
both with and without the zero point energy (ZPE) and the RT value: the differences among them are
of the order of hundredths of angstroms. The computed RT $a_0$ is in good agreement
with the RT experiment~\cite{Kang_2019} (the difference is smaller
than $\approx$ 0.02 \AA).
The comparison with other computed $a_0$ is also reported: the LDA values agree within $\approx$ 1 \%. This small differences depend
primarily on the computational parameters chosen for the calculation and from
the different pseudopotentials.
The PBE values of $a_0$ are slightly larger than the LDA ones as usually found with the
functionals that use the generalized gradient approximation (GGA).

\begin{figure}
\centering
\includegraphics[width=\linewidth]{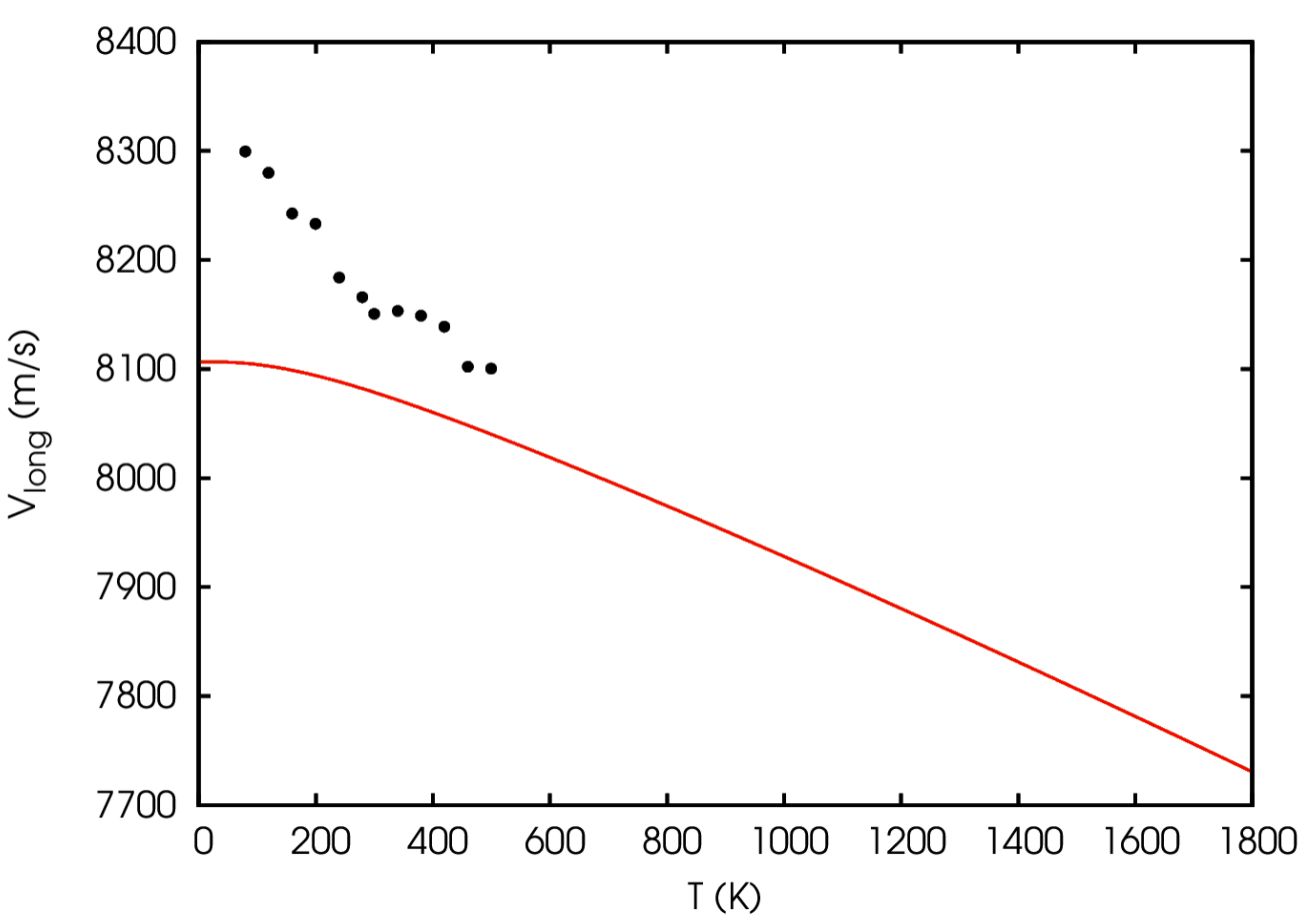}
\caption{Longitudinal sound velocity in the [111] direction: experimental points
(extrapolated from Kang {\it et al.}~\cite{Kang_2019}) and this work (red curve).}
\label{fig_v}
\end{figure}

In Table~\ref{table1} we report the calculated values of the elastic constants
together with other theoretical estimates available in the literature and
the experimental values. The LDA values are closer to experiment, while
the PBE (Perdew Burke Ernzerhof) calculations give smaller ECs, as usually found 
with the GGA functionals.
At $T=0$ K, the softening due to the ZPE is $\approx 2.4 \%$ for $C_{11}$,
$\approx 1.2 \%$ for $C_{12}$ and $\approx 2.4 \%$ in $C_{44}$.
In our calculation we relax the ionic positions for each strained configuration
so the ECs are relaxed-ions results. This is relevant only
for $C_{44}$ while for $C_{11}$ and $C_{12}$ the ionic positions are determined
by symmetry. The frozen-ions results, obtained by a uniform strain of the
atomic positions but no further relaxation, are also shown in Tab.~\ref{table1}:
$C_{44}$ is $\approx$ 14 \% larger than the relaxed-ions value.
Our theoretical $T=300$ K values are in good agreement with experiment:
$C_{11}$ is smaller of $\approx$ 1.5 \%, $C_{12}$ is smaller of $\approx$ 5.1 \%
and $C_{44}$ is larger of $\approx$ 1 \%.

\begin{figure*}
\centering
\includegraphics[width=.49\linewidth]{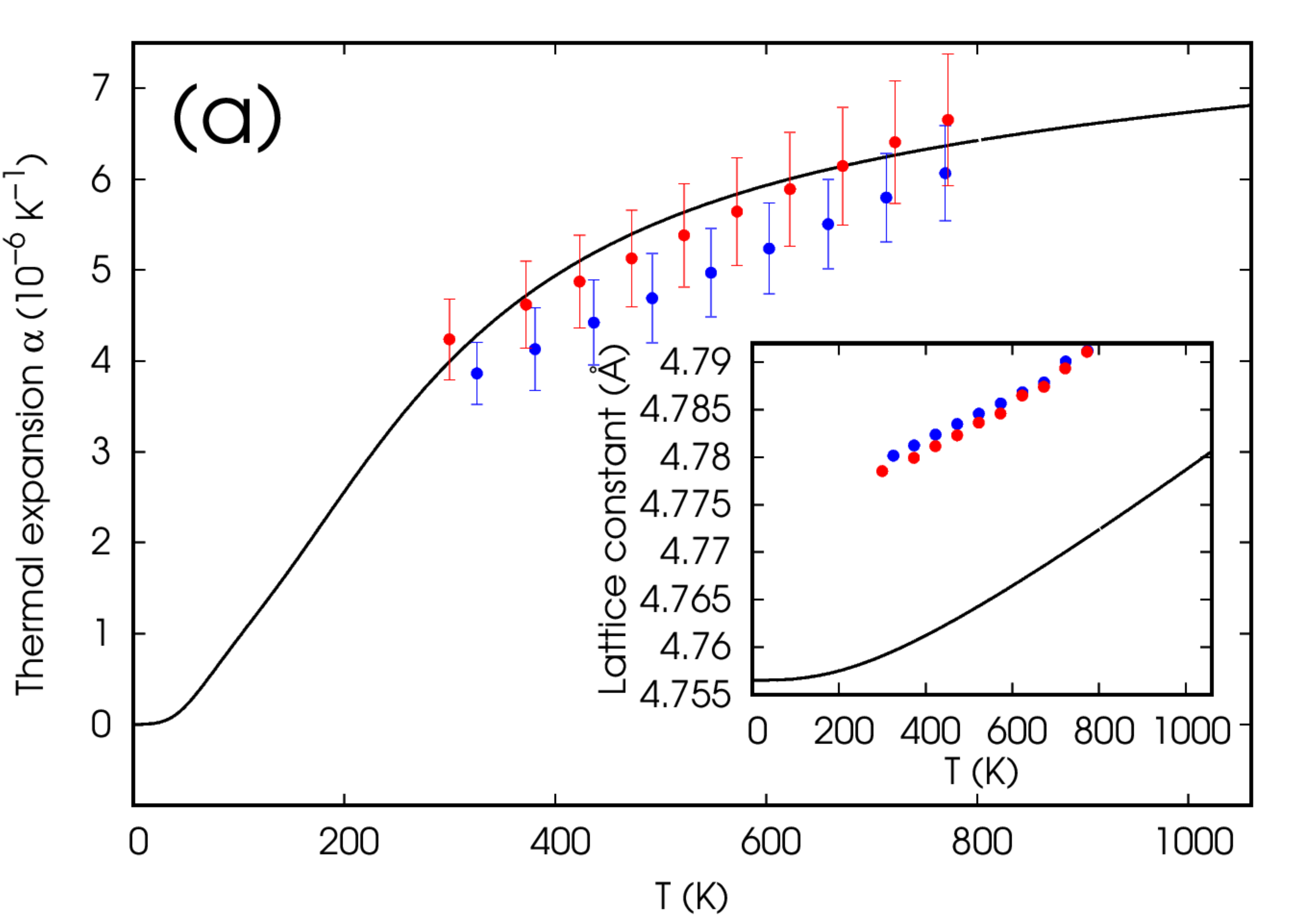}
\includegraphics[width=.49\linewidth]{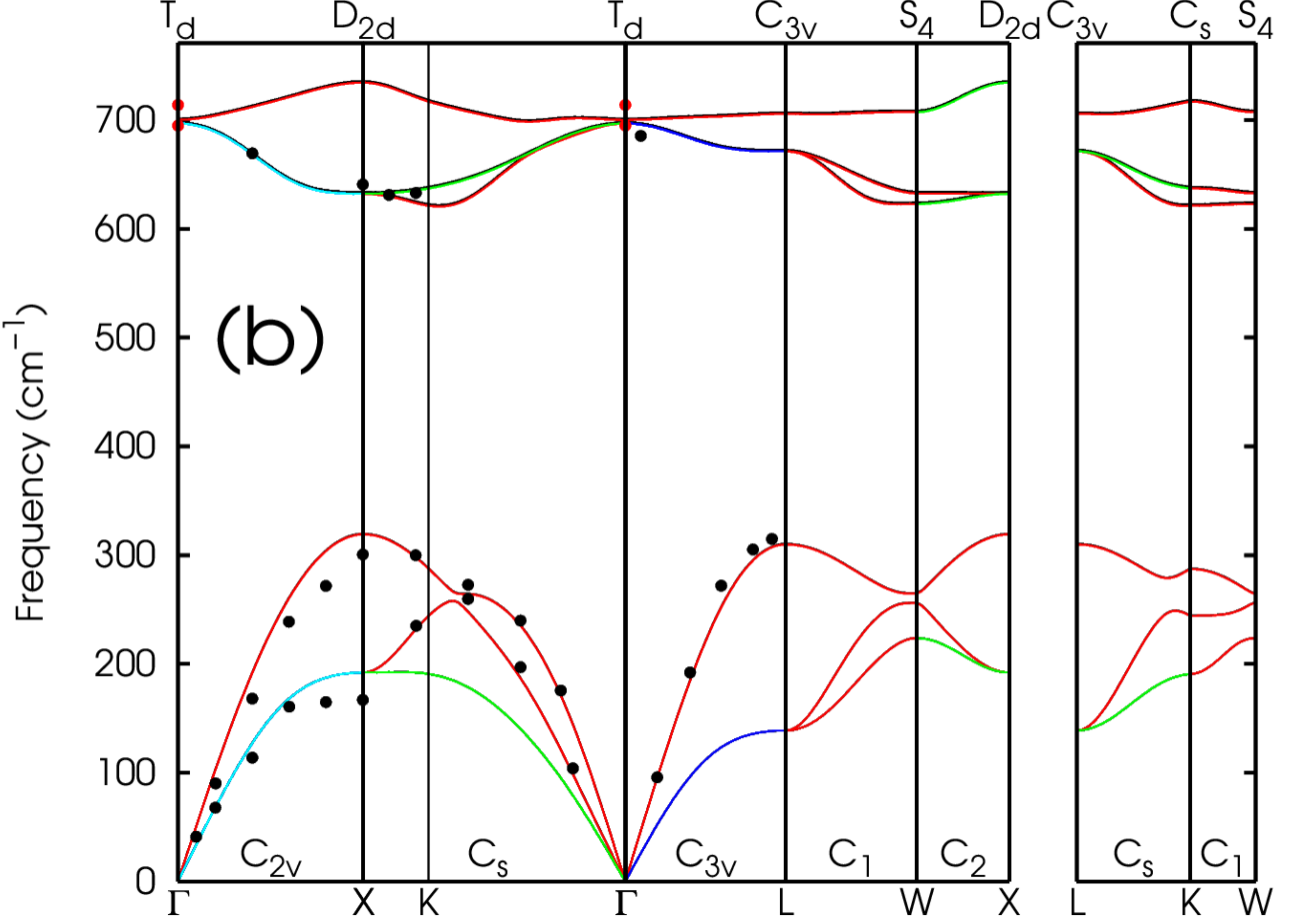}
\includegraphics[width=.49\linewidth]{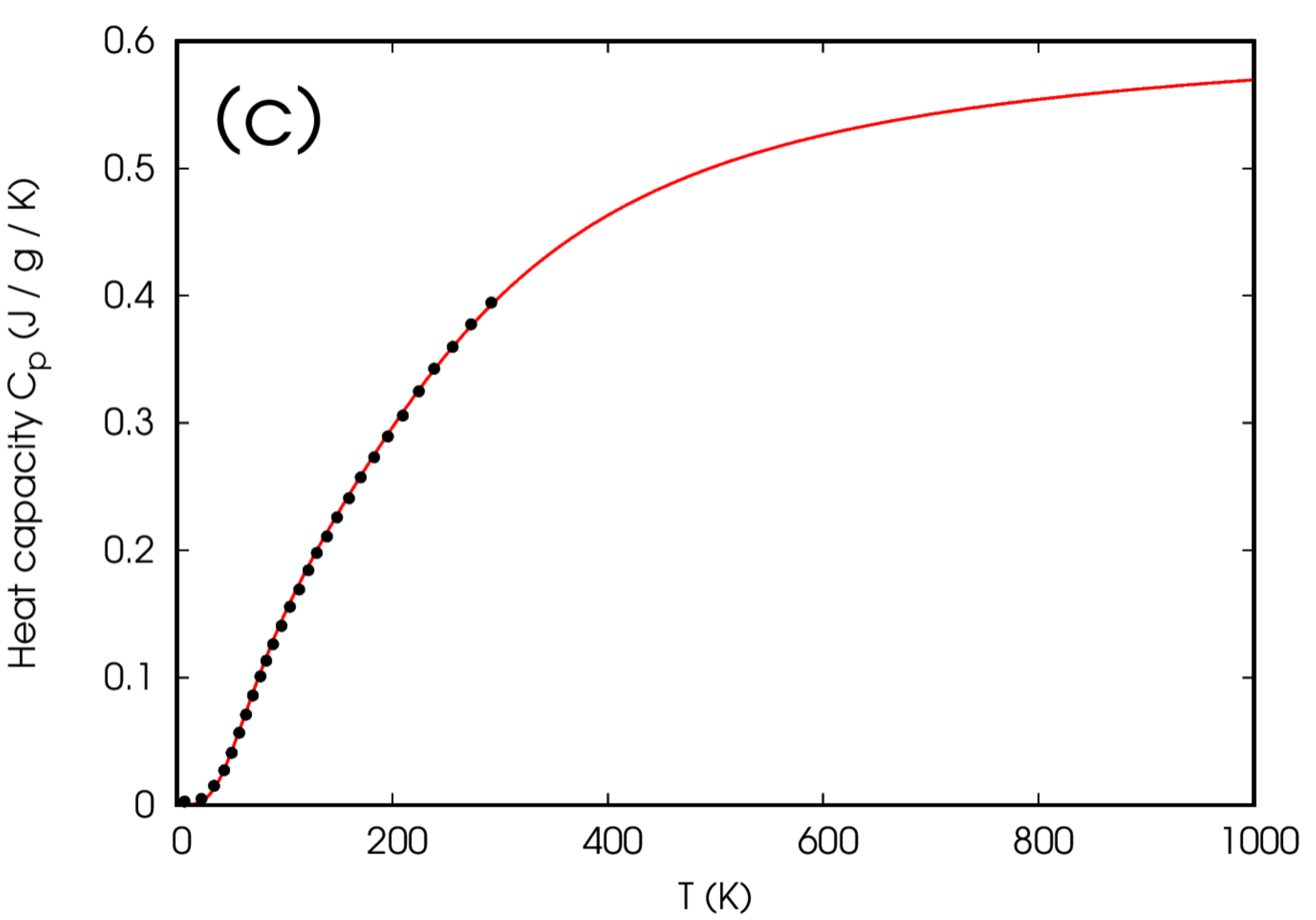}
\includegraphics[width=.49\linewidth]{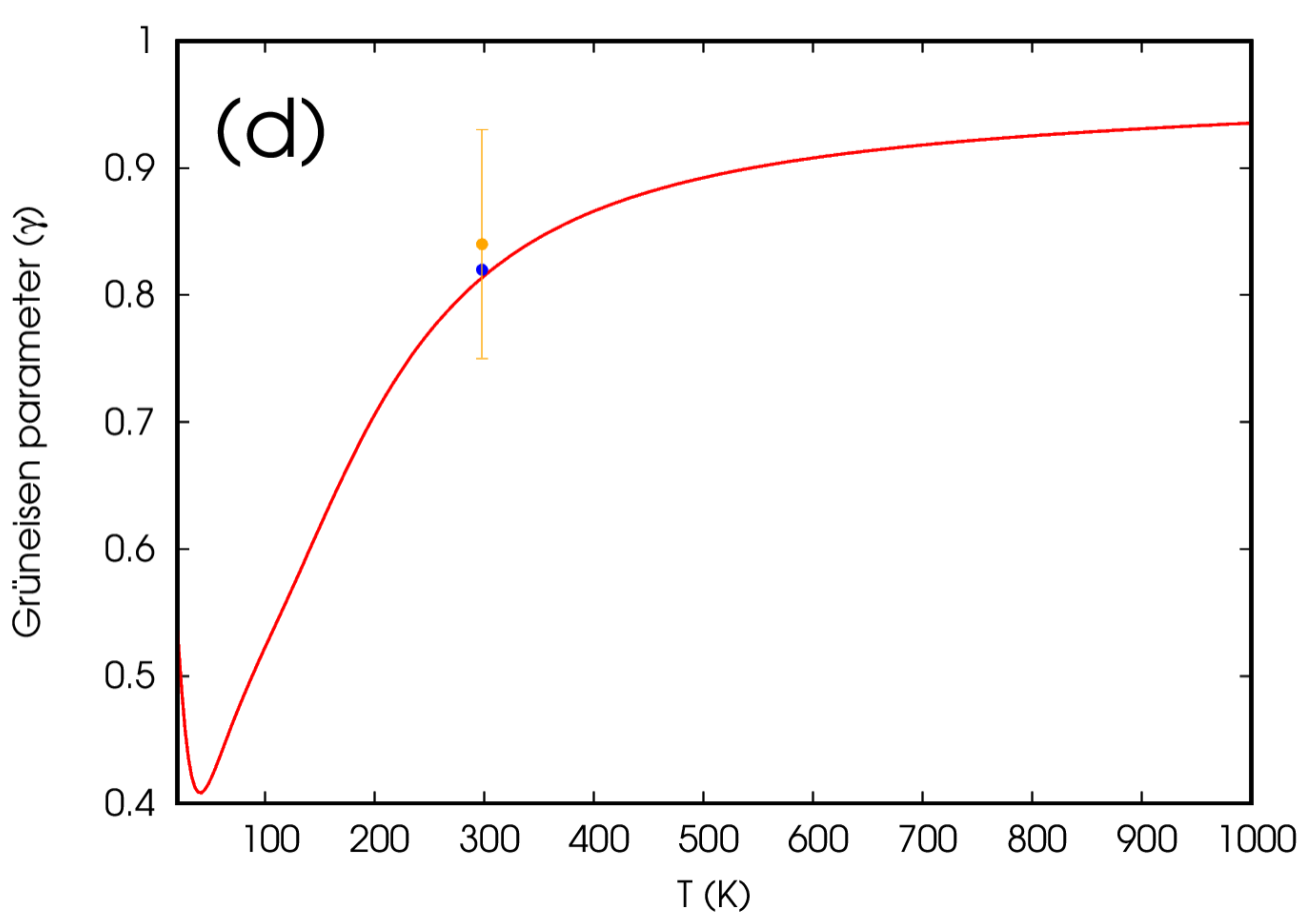}
\caption{(a) Thermal expansion and temperature dependence of the lattice parameter (inset). Black lines represent this work,
        the points are experimental data: blue~\cite{Kang_2019} and red~\cite{Chen_2019}.
        (b) Phonon dispersion curves at $T=300$ K (colored lines) and experimental points: black \cite{Ma_2016} and red \cite{Greene_1994}.
            The colours of the phonon branches indicate their symmetry (as explained in the thermo\_pw documentation).
            Phonons curves at $T=0$ K (black lines, barely visible only in the optical branches).
        (c) Computed isobaric heat capacity (red line) and comparison with experiment (points) adapted from Chen {\it et al.}~\cite{Chen_2019}.
        (d) Average Gr\"{u}neisen parameter. This work (red curve) and experimental points: blue~\cite{Kang_2019} and orange~\cite{Chen_2019}.}
\label{fig_te}
\end{figure*}

In Figure~\ref{fig_ec} the TDECs and the bulk modulus are reported.
The red lines are the adiabatic ECs (or bulk modulus), while
the blue lines are the isothermal ECs.
Evaluating the percentage softening of ECs due to temperature as:
\begin{equation}
{{C_{ij}(T=0 {\rm K}) - C_{ij}(T=1800 {\rm K})} \over {C_{ij}(T=0{\rm K})}}\times 100, 
\end{equation}
where the ZPE is included in $C_{ij}(T=0 {\rm K})$ we find:
$\approx$ 11 \% (adiabatic) and $\approx$ 13 \% (isothermal) for $C_{11}$,
$\approx$ 9 \% (adiabatic) and $\approx$ 15 \% (isothermal) for $C_{12}$
and $\approx$ 13 \% for $C_{44}$.

As far as we know, presently no experimental measurement of the TDECs is available
to compare directly with our result, but recently the temperature
dependence of the longitudinal sound velocity along the $[111]$ direction was measured
till $\approx 500$ K~\cite{Kang_2019}. Using the adiabatic TDECs and the density $\rho$ this
sound velocity can be written as~\cite{kittel}:
\begin{equation}\label{vlong}
V_{long}=\left(  \frac{C_{11} + 2 C_{12}+ 4 C_{44}}{3 \rho} \right)^{\frac{1}{2}}
\end{equation}
and in Fig.~\ref{fig_v} we compare our theoretical
estimate with experiment. We take into account the temperature dependence of the
density due to TE effect. The use of a temperature independent density in
Eq.~\ref{vlong} (for instance the density at $T=0$ K or $T=300$ K) leads to an
appreciably lower sound velocity above the RT (for instance
at $T=1500$ K the difference is $\approx$ 90 m/s).
The theoretical values of the sound velocity as a function of temperature is
in reasonable agreement with experiment in the analyzed range, but the
experimental slope of the curve at low temperatures is not reproduced. The comparison improves above RT. The difference between theory and experiment at RT ($0.9 \%$) is compatible with the errors in the elastic constants and in the density, while the difference at the lowest measured temperature ($2.3 \%$) is larger. 


We can also compare the TDECs of BAs with those of silicon that we have
calculated by using LDA~\cite{Malica_2020}. Silicon ECs are
smaller, the $T=0$ K values (with ZPE) are: $C_{11}=1580$ kbar, $C_{12}=639$ kbar and
$C_{44}=746$ kbar. In the range of temperature calculated for silicon ($0-800$ K)
$C_{11}$ and $C_{12}$ decrease of about $7$ \% and $C_{44}$ of $5$ \%
(for both isothermal and adiabatic).
In the same range of temperature the decrease of the ECs of BAs is slightly smaller:
$3.8$ \% (adiabatic) and $4.5$ \% (isothermal) for $C_{11}$,
$3.2$ \% (adiabatic) and $5.7$ \% (isothermal) for $C_{12}$ and $4.5$ \% for $C_{44}$.

We now present several other thermodynamic properties of BAs that have been
calculated together with the TDECs. The TE and the temperature dependence
of the lattice constant $a(T)$ are shown in Figure~\ref{fig_te} in the range of
temperature $0-1100$ K. We compare our numerical result (black line) with two
recent measurements in the temperature range $300-773$ K (red points)~\cite{Chen_2019} and
$321-773$ K (blue points)~\cite{Kang_2019}. The experimental $a(T)$ is higher
than the theoretical one by approximately $0.02$ \AA, but the temperature dependence
is reproduced correctly as visible in the TE plot reported in the same figure.
In the experimental temperature range, the experimental $a(T)$ are both linear
with $T$ but the slope
of the red points is slightly larger and more in agreement with our calculation.
This behavior is reflected in the TE: the TE
of Chen {\it et al.}~\cite{Chen_2019} agrees with ours within experimental uncertainties while the
TE of Kang {\it et al.}~\cite{Kang_2019} has slightly smaller values (although within the error bar of the former~\cite{Chen_2019}).
In particular the RT TEs are $4.0 \times 10^{-6} K^{-1}$ (this work),
$(4.2 \pm 0.4) \times 10^{-6} K^{-1}$ (Chen {\it et al.}~\cite{Chen_2019}) 
and $3.85 \times 10^{-6} K^{-1}$ (Kang {\it et al.}~\cite{Kang_2019}).
Other DFT-LDA estimates are: $4.0 \times 10^{-6} K^{-1}$
(Chen {\it et al.}~\cite{Chen_2019}), and $3.04 \times 10^{-6} K^{-1}$ (Broido {\it et al.}~\cite{Broido_2013}).
Molecular dynamic simulations~\cite{Benkabou_1999} produced the result $4.1 \times 10^{-6}$.
A much larger value has appeared recently~\cite{Daoud_2019} $10.9 \times 10^{-6} K^{-1}$
by using DFT method within the GGA, quite far from experiment.

The phonon frequencies computed at the different geometries are interpolated at
$T=300$ K and the result is shown in Figure~\ref{fig_te}b (colored lines).
The phonons are compared with RT measurements obtained from inelastic X-ray
scattering \cite{Ma_2016} and Raman spectroscopy \cite{Greene_1994} (points). As already
found in previous references~\cite{Broido_2013, Ma_2016} the agreement between theory and experiment is quite good.
The difference between RT and $T=0$ K phonons (black lines) is only barely visible
in the optical branches.

The isobaric heat capacity is shown in Figure~\ref{fig_te}c. The points indicate the temperature dependence of experimental data below RT \cite{Chen_2019} that we extrapolated from the plot.
The agreement is very good, consistently with the fact that also the calculated Debye temperature ($\Theta_D=668$ K, obtained from the
$T=0$ K+ ZPE ECs) is in good agreement with the experimental
value~\cite{Chen_2019} ($\Theta_D=681$ K).



The temperature dependence of the average Gr\"{u}neisen parameter is reported in Figure~\ref{fig_te}d. 
We also report the RT experimental values of Kang {\it et al.}~\cite{Kang_2019} (blue point) and Chen {\it et al.}~\cite{Chen_2019} (orange
point). The experimental uncertainty of the first point is not known, but
it is very close to our curve. The second point is in agreement with our estimate
within the experimental error bar.

\begin{figure}
\centering
\includegraphics[width=\linewidth]{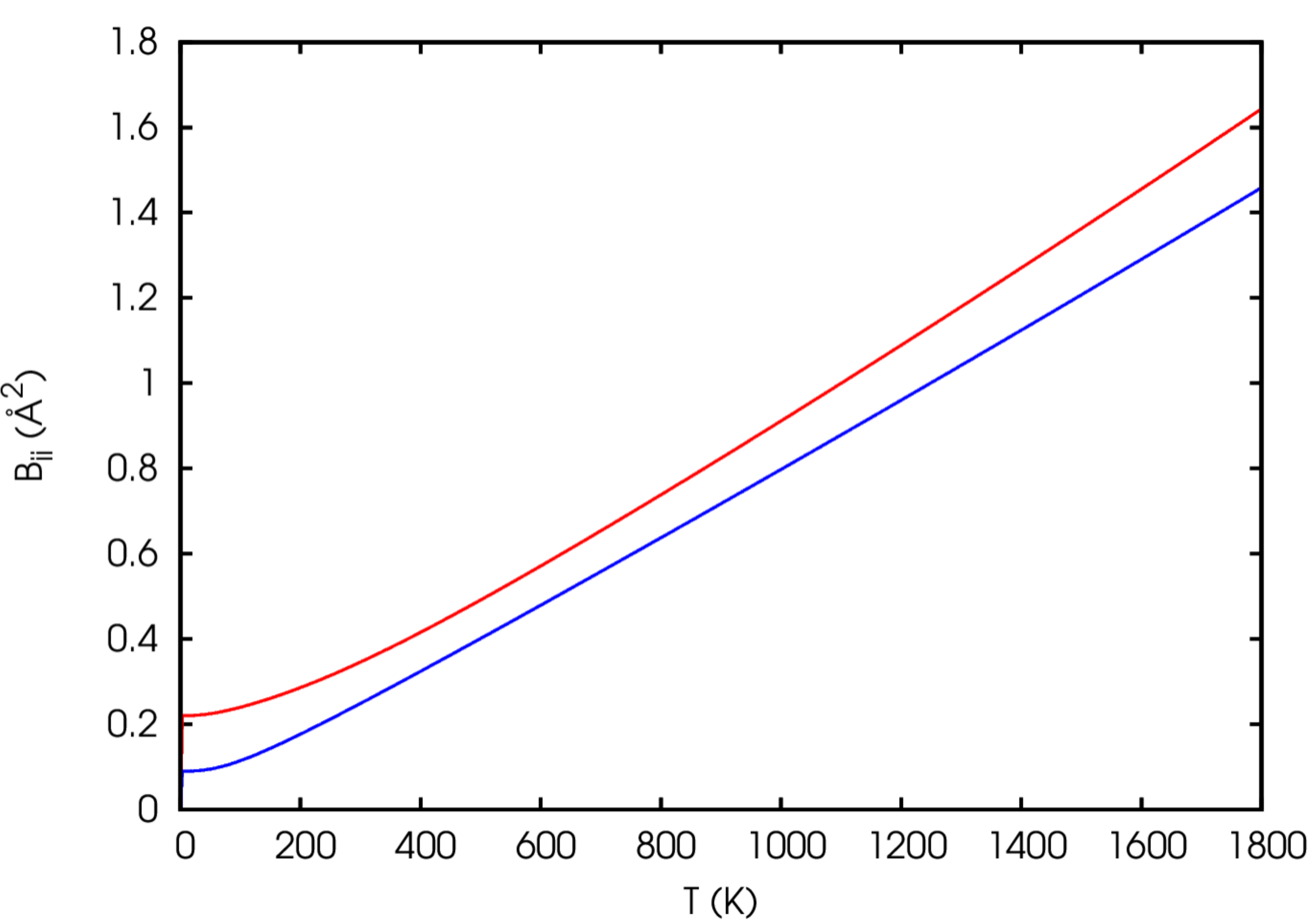}
\caption{Temperature dependent atomic B-factor for the cubic BAs: boron (red), arsenic (blue).}
\label{fig_dwf}
\end{figure}

Finally in Figure~\ref{fig_dwf} we show our estimate of the BF for boron (red)
and arsenic (blue). The BF of boron is 0.35 \AA$^2$ at $T=300$ K and grows
up to 0.74 \AA$^2$ at $T=800$ K. The BF of arsenic is smaller because of the
larger mass: 0.25 \AA$^2$ at $T=300$ K and 0.64 \AA$^2$ at $T=800$ K.
They are smaller than the
BFs of silicon~\cite{Malica_2019} which has 0.52 \AA$^2$ at $T=300$ K and 1.29 \AA$^2$ at
$T=800$ K.

\section{Conclusions}
We studied TDECs of BAs by means of ab-initio simulation by using
the recent tools implemented in the \texttt{thermo\_pw} code within
the QHA. The values of the adiabatic ECs at $T=300$ K are
$C_{11}=2807$ kbar, $C_{12}=754$ kbar,
$C_{44}=1507$ kbar. We found that in the
range of temperature $0-1800$ K the percentage softening of the adiabatic ECs is
$\approx$ 11 \% for $C_{11}$, $\approx$ 9 \% for $C_{12}$ and $\approx$ 13 \%
for $C_{44}$. The order of magnitude of the softening is consistent with
the temperature variation of the longitudinal sound velocity measured along the
$[111]$ direction~\cite{Kang_2019}. The slope of the curves are different below
RT and similar for larger temperatures.
The softening computed in BAs is slightly smaller than that
calculated for silicon in the temperature range $0-800$ K~\cite{Malica_2020}.
We also computed thermodynamic properties of BAs such as the TE, the heat
capacity, and the average Gr\"uneisen parameter finding good agreement
with experiments. Finally we presented
the calculated atomic BFs as a function of the temperature for which no
information, neither experimental nor theoretical, was available so far.

\begin{acknowledgments}
Computational facilities have been provided by SISSA through its Linux
Cluster and ITCS and by CINECA through the SISSA-CINECA 2019-2020
Agreement.
\end{acknowledgments}


\nocite{*}
\bibliography{aipsamp}

\end{document}